# Generation and manipulation of multipole and vortex events in (1+1)-dimensional spacetime

Yiming Pan[1], Guowei Chen[1]

[1]State Key Laboratory of Quantum Functional Materials, School of Physical Science and Technology and Center for Transformative Science, ShanghaiTech University, Shanghai 200031, China

**Abstract**

An event is usually regarded as a zero-dimensional coordinate label in spacetime, rather than as an object with spatial extent, internal structure or topology. Here we show that in nonlinear photonic spacetime crystals, a complete energy-momentum gap ($\omega k$-gap) suppresses extended radiation channels, while Kerr self-trapping localizes an optical occurrence in the physical $(x, t)$ plane. Imposing and relaxing symmetry constraints then yields multipole and vortex events with distinct internal structure. Their spacetime action of these states establishes a hierarchy of structural cost, Bogoliubov-de Gennes spectra distinguish phase-winding vortices from real-valued multipoles, and initial-value reconstructions provide an independent robustness check together with a finite preparation window. Using representative optical parameters, we predict that these structured events span transverse scales of 30–60 μm and temporal duration of 85–170 fs. These results suggest that an event can be treated not only as a point label, but also as a localized wave object with controllable internal degrees of freedom, offering a route to wave control in time-varying nonlinear media.

In relativity, an event denotes an elementary occurrence in spacetime: a position and a time. It is normally a coordinate label, rather than a physical object, a propagating pulse or a state with internal degrees of freedom. However, point-like idealizations can acquire effective size and structure when embedded in a wave medium: wave packets represent particles, light pulses carry temporal profiles, and optical solitons may host multipole or vortex structures [1]. This observation motivates our central question: can a finite event in a wave medium behave as an object in its own right, with symmetry, topology and internal structure?

Photonic spacetime crystals (STCs) provide a suitable setting, as their refractive index is modulated in both space and time [2–6]. Such media can open combined frequency-momentum gaps, or complete $\omega k$-gaps [7–10]. Recent studies have also shown that event-like wave packets can be trapped by topological spacetime defects [8,11] or by Kerr nonlinearities [8,12]. Here we focus on a different form of localization: an object confined in real space and real time, rather than in a propagation coordinate and a retarded time. Its lobes, nodes and phase winding are therefore internal structures of a finite occurrence in the physical $(x, t)$ plane, rather than transverse beam textures or conventional spatiotemporal wavepacket features. In this sense, these event-molecule structures differ from spatial solitons [13–15], temporal solitons [12,16], light bullets [17–19] 10 and spatiotemporal wave packets [20,21].

Inside a complete $\omega k$-gap of nonlinear STCs, the dimensionless envelope equation can be written as $(1 - \partial_{tt})(1 - \partial_{xx})\psi - 4|\psi|^2\psi = 0$ [8]. Its linear spectrum contains a semi-infinite gap below the continuum, while the Kerr term provides the attractive nonlinearity required for self-trapping (Fig. 1a). The solution $\psi = sech(x)sech(t)$ represents the elementary event state: a frozen and transient occurrence in the STC medium, rather than a pulse evolving through it. The aim here is to give this elementary event higher-order spacetime structure. Although multipoles and vortices are familiar signatures of internal organization in two-dimensional nonlinear systems [1], in the present setting they are embedded in the event itself, with spacetime lobes and phase singularities bound by nonlinear self-trapping and organized by the spacetime symmetry.

We construct multipole and vortex event excitations in nonlinear photonic STCs as a hierarchy of structured localized states. Our analysis proceeds in three steps. First, we define a spacetime action to identify finite events and quantify the cost of their internal multipole or vortex structure. Second, we use Bogoliubov-de Gennes (BdG) spectroscopy to resolve their bound deformation channels. Third, we use initial-value simulations as a dynamical reconstruction test: starting from nearby slices of the stationary profile, we determine whether the target core can be recovered within a finite preparation window before unstable directions dominate. For the representative parameters used here (Methods 1), one dimensionless unit corresponds to about 6.37 $\mu m$ and 17.19 $fs$, given event extent of about 30–60 $\mu m$ and 85–170 $fs$. The mixed spacetime operator reduces continuous rotational symmetry to the discrete $D_4$ symmetry, supporting real nodal-line multipoles and complex vortices with phase winding around a spacetime singularity. This places the finite event, rather than a propagating pulse, at the center of wave design in a time-varying medium.

**Model and spacetime action**

A nonlinear photonic spacetime crystal with complete $\omega k$-gap provides a minimal setting for finite event localization (Methods 1). In dimensionless units, the event field $\psi(x,t)$ obeys

$$(1 - \partial_{tt})(1 - \partial_{xx})\psi - 4|\psi|^2\psi = 0. \quad (1)$$

Here $x$ and $t$ are coordinates of the event, rather than a propagation coordinate and a retarded time. The linear operator describes the combined opening of momentum and frequency gaps, while the cubic term represents the self-focusing Kerr response. For a plane wave $exp(ik_x x - ik_t t)$, the linear spectrum is

$$\lambda(k_x, k_t) = (1 + k_x^2)(1 + k_t^2) \geq 1. \quad (2)$$

The linear continuum therefore starts at $\lambda = 1$, leaving a semi-infinite gap ($\omega k$-gap) below it. A localized state in this gap has no resonant extended linear radiation channel. Kerr self-focusing can then support localization in both space and time. The fundamental solution $\psi(x,t) = sech(x)sech(t)$ defines the elementary event state, or monopole event. Figure 1a summarizes this mechanism: a plane-wave excitation belongs to the linear continuum; the complete $\omega k$-gap suppresses extended radiation channels; and Kerr attraction supports a localized spacetime occurrence.

Equation (1) should be treated as a spacetime boundary-value problem, rather than as a standard initial-value problem. Since the physical time coordinate is part of the event itself, a generic initial slice can excite evanescent components and is highly sensitive to the prescribed field and derivative data. We therefore construct complete spacetime profiles as stationary points in function space. The corresponding spacetime action is

$$S[\psi] = \int ( |\psi|^2 + |\partial_x\psi|^2 + |\partial_t\psi|^2 + |\partial_x\partial_t\psi|^2 - 2|\psi|^4) dx\, dt. \quad (3)$$

The Euler-Lagrange condition ($\delta S/\delta\psi^* = 0$) gives Eq. (1). This action defines a variational landscape for event states and event molecules: each localized solution is a stationary point, while its spacetime lobes, nodal lines or phase singularities determine the corresponding sector of function space. The mixed-derivative term gives the functional its spacetime character. By coupling $x$ and $t$, it reduces continuous rotational symmetry to the discrete $D_4$ symmetry of the $(x,t)$ plane (Method 2). This symmetry permits multipole events with nodal lines and vortex events with phase winding around a spacetime singularity.

We use the variational construction to estimate the relevant action scale. For the ansatz $\psi(x,t) = A\, f(x,t;\alpha)$, the action becomes $S(A,\alpha) = A^2L(\alpha) - A^4N(\alpha)$. Optimizing over the amplitude gives $A^2 = \frac{L(\alpha)}{2N(\alpha)}$ and $S_{red}(\alpha) = \frac{L(\alpha)^2}{4N(\alpha)}$. The reduced action ranks candidate event molecules and provides symmetry-compatible initial guesses. The final profiles are obtained by symmetry-projected Petviashvili relaxation (Method 3). During relaxation, the field is allowed to deform over the full $(x, t)$ plane, while parity and phase-winding projections prevent higher-order states from relaxing into the monopole. This procedure yields the monopole, dipole, vortex, quadrupole and octupole event profiles shown in Fig. 1b.

Figure 1 demonstrates a hierarchy of structured event states, rather than merely presenting a set of numerical solutions. The monopole is the elementary event state: a single self-trapped occurrence localized in both $x$ and $t$. The dipole is the simplest two-lobed event molecule and has temporal and spatial variants. Vortex, quadrupole and octupole states contain multiple spacetime profiles organized by phase singularities or nodal constraints. For the stationary solutions considered here, the actions follow the hierarchy $S_{monopole} < S_{dipole} < S_{vortex} < S_{quadrupole} < S_{octupole}$, indicating the increasing cost of encoding internal spacetime structure within a finite event. The variational estimates and the stationary values obtained from iterative relaxation are compared in Supplementary Fig. S1, supporting the consistency of this structural-cost hierarchy.

In function space, different multipole orders correspond to different regions of the spacetime-action landscape. The action $S[\psi]$ contains spatial and temporal gradient contributions, including $|\partial_x \psi|^2$ and $|\partial_t \psi|^2$. A higher-order multipole introduces sign-changing nodal lines in the spacetime. Across these lines, the field changes from positive to negative amplitude, increasing the gradient contribution to the action. The monopole event is therefore the lowest nontrivial stationary state, whereas higher-order multipole events appear as excited saddle states whose action is raised by their nodal structure.

The vortex event has a different action cost because nonlinear binding also contributes to its formation. At the level of linear superposition, it can be viewed as a quadrature pair formed from a spatial dipole and a temporal dipole, for which the linear part of the action is approximately additive. The distinction arises from the self-focusing nonlinear term, $N \propto \int |\psi|^4 \, dxdt$. For a dipolar ansatz of the form $(x + it)\, exp[-\alpha(x^2 + t^2)]$, $|\psi|^4$ contains $x^4 + t^4 + 2x^2t^2$. The cross term $2x^2t^2$ represents nonlinear overlap between the spatial and temporal dipoles, analogous to a cross-phase-modulation contribution. It increases $N$ and therefore lowers the reduced action $S_{red} \propto L^2/N$. In this sense, the spacetime vortex may be interpreted as a bound state of two dipolar components sharing a self-induced potential well, which explains why its action can be lower than that of a real quadrupole with extended nodal lines.

The hierarchy also distinguishes two types of internal spacetime topology. A real quadrupole retains nodal lines along both spacetime axes, which produces gradient costs over extended regions of the $(x, t)$ plane. By contrast, a vortex concentrates its topological defect at a phase singularity, while the surrounding phase varies continuously. This point-defect structure helps explain why the vortex can have a lower action than the quadrupole, despite their similar four-sector intensity patterns. In this sense, the action hierarchy is not only a numerical ordering, but also reflects how event molecules encode internal spacetime structure.

## BdG spectroscopy of event molecules

After obtaining the stationary event profiles, we examine how each event molecule can deform in spacetime. This is not a conventional dynamical-stability problem, because $t$ is itself a coordinate of the event. We therefore probe the local curvature of the spacetime action by expanding around a stationary profile $\psi_0(x,t)$. Writing $\psi = \psi_0 + \delta\psi$, the first-order variation vanishes, and the leading correction is quadratic:

$$S[\psi_0 + \delta\psi] = S[\psi_0] + \left(\frac{1}{2}\right)\int (\delta\psi^*, \delta\psi)\mathcal{H}\,(\delta\psi, \delta\psi^*)^T dx\,dt + \text{higher-order terms.} \qquad (4)$$

For a complex vortex event, amplitude and phase perturbations are coupled by the background phase winding. In the Nambu basis $(\delta\psi, \delta\psi^*)^T$, the second-variation operator $\mathcal{H}$ takes the Bogoliubov-de-Gennes (BdG) form

$$\mathcal{H} = \begin{bmatrix} L_{xt} - 8|\psi_0|^2 & -4{\psi_0}^2 \\ -4(\psi_0^*)^2 & L_{xt} - 8|\psi_0|^2 \end{bmatrix}, \qquad (5)$$

where $L_{xt} = (1 - \partial_{tt})(1 - \partial_{xx})$ is the linear spacetime operator term (Method 4). We solve $\mathcal{H}v_n = \lambda_n v_n$, where $\lambda_n$ gives the local curvature of the action landscape and $v_n(x,t)$ gives the associated spacetime deformation channel. For real-valued multipole events, the amplitude and phase fields decouple, the BdG operator then reduces to $\mathcal{H}_{real} = L_{xt} - 12|\psi_0|^2$.

The BdG spectrum separates localized bound modes from extended continuum fluctuations. Far from the event, $\psi_0$ vanishes and the operator approaches the vacuum spectrum $\lambda(k_x, k_t) = (1 + k_x^2)(1 + k_t^2)$, whose continuum starts at $\lambda = 1$. Modes below this threshold ($\lambda < 1$) are confined by the nonlinear potential and correspond to internal deformation channels, whereas modes above the threshold represent extended radiation-like fluctuations.

Figure 2 presents the BdG spectra of the vortex and quadrupole events. In both cases, discrete modes below $\lambda = 1$ are localized near the event core, while modes above the threshold extend across the spacetime domain. These bound modes provide spectral fingerprints of the event molecules, indicating deformation channels associated with breathing, splitting, phase rotation or motion toward lower-action configurations. The corresponding spectra of the monopole, spatial dipole, temporal dipole, and octupole events are provided in Supplementary Figs. S2–S5, together with their bound deformation and radiation channels.

The vortex-quadrupole comparison shows that similar four-sector intensity patterns need not imply the same internal topology. Bound modes of the vortex inherit the point phase defect of the background, $\psi_0 \sim (x + it)$, and therefore couple amplitude and phase deformations. By contrast, quadrupole modes are mainly real parity-changing deformations associated with nodal lines, $\psi_0 \sim xt$. The BdG spectrum thus provides a way to distinguish a vortex event from a quadrupole event even when their intensity distributions appear similar.

In an unconstrained physical system, the monopole is expected to be the most structurally robust event state. Dipoles, quadrupoles, octupoles and vortex events are higher stationary states maintained by symmetry, topology or preparation conditions, and are therefore generally unstable when the field can explore the full action landscape. This instability is not a numerical artefact, but reflects the saddle-point character of these solutions in $S[\psi]$. Because $t$ is localized together with $x$, the instability should not be interpreted as the slow decay of a beam during laboratory propagation. Rather, it describes a transient nonlinear reorganization of energy in spacetime, closer to a resonant scattering event than to a long-lived soliton beam.

Because higher-order events are saddle points rather than minima of the action, they are expected to possess of low or negative-curvature. These directions do not preclude the existence of stationary event states. Instead, they identify deformation channels that render their preparation more demanding. The BdG spectrum therefore connects the variational existence of these states with the independent dynamical reconstruction test introduced below.

We compute the low-lying localized spectrum using sparse shift-and-invert Arnoldi iteration (Method 4). Modes near the continuum edge may depend on box size, grid resolution and spectral shift. We therefore assess localization and convergence together with eigenvalue position to distinguish genuine bound modes from finite-domain artefacts. In this way, Fig. 2 provides the spectral map used in the following initial-value simulations, identifying the deformation channels that control transient assembly and breakdown.

**Initial-value reconstruction within a finite spacetime window**

The action and BdG analyses characterize the stationary event profiles and their local deformation channels. The remaining question is whether these solutions can be recovered dynamically from nearby initial data, providing a robustness check independent of the symmetry-projected Petviashvili construction.

We address this by using each boundary-value solution to define a controlled initial-value problem (IVP). For a target event $\psi_{target}(x,t)$, we choose a negative-time slice $t_0$ before the strongest nonlinear interaction, and prescribe both $\psi_{target}(x,t_0)$ and $\partial_t \psi_{target}(x,t_0)$. The derivative is obtained from the full target profile, reducing finite-difference errors that could seed evanescent components. This benchmark is not intended as an optimized experimental launch protocol. Rather, it tests whether a nearby dynamical trajectory can reconstruct the target core without imposing the complete stationary profile. The evolution is integrated using a Fourier pseudospectral spatial discretization and a fourth-order Runge-Kutta time step (Method 5).

Figure 3 applies this reconstruction test to the vortex event. The target profile contains a finite-amplitude core carrying an $m = 1$ spacetime phase winding. When the initial slice is chosen sufficiently close to the core, the IVP evolution reconstructs both the intensity profile and four-sector phase pattern near $t \approx 0$. The vortex is therefore not merely a stationary solution by symmetry-projected relaxation, but can be dynamically recoverable within a finite preparation window. After formation, however, the trajectory leaves the local neighborhood of the stationary state and develops blow-up accompanied by phase dislocations. Choosing an earlier initial slice increases the pre-assembly evolution time, allowing small spectral and truncation errors to grow and disrupt the phase winding. The contrast between the two initial slices therefore defines a finite preparation window associated with the unstable deformation spectrum.

Figure 4 extends the same test to real higher-order multipoles. The quadrupole can be transiently reconstructed from a sufficiently late initial slice, but its nodal-line structure subsequently develops fingering and blow-up after the core interaction. The octupole is more fragile: its additional lobes and sign changes require a broader, lower-amplitude initial pattern, so small mismatches are rapidly amplified into filamentary structures. This order-dependent fragility is consistent with the action hierarchy in Fig. 1b and with the increasing number of bound deformation channels identified by the BdG spectrum.

Together, Figs. 3 and 4 distinguish three questions that can otherwise be conflated. Existence means that a structured event is a stationary point of the spacetime action. Spectral structure means that the BdG operator contains bound deformation channels rather than only continuum radiation. Dynamical recoverability further requires that nearby initial data remain close enough to the target profile for the core to reconstruct within a finite window. The higher-order states satisfy the first condition, show the second through bound modes, and satisfy the third only over relatively narrow windows. For comparison, Supplementary Fig. S6 shows the corresponding IVP evolutions for the monopole and dipole states, which are structurally more persistent.

**Further discussion**

The central practical issue is not indefinite stability, but reconstruction within a finite preparation window. Multipole and vortex events are saddle points of the spacetime action, and generic initial-value trajectories are expected to move away from them. Their physical relevance depends on whether a near-target input can approach the target profile and reconstruct its core before unstable directions dominate.

The observed blow-up, spacetime azimuthal breakup and phase dislocations should therefore be interpreted as signatures of a limited preparation window, rather than as evidence against the existence of the stationary states. In this sense, the IVP analysis plays a role complementary to the symmetry-projected Petviashvili solver: it provides an independent dynamical robustness check on the localized states identified by the boundary-value problem. The action hierarchy estimates the cost of internal spacetime structure, the BdG spectrum identifies deformation channels, and the IVP test measures how accurately and how long a nearby trajectory can reconstruct the target core.

We also emphasize that the vortex states are constrained by the spacetime symmetry of the action. The linear part $L_{xt} = (1 - \partial_{tt})(1 - \partial_{xx}) = 1 - \partial_{xx} - \partial_{tt} + \partial_{xx}\partial_{tt}$ contains the mixed term $\partial_{xx}\partial_{tt}$, which breaks continuous rotations in the $(x, t)$ plane to the discrete square $D_4$ symmetry. A circular vortex is therefore not an eigenstructure naturally selected by this operator; instead, the $D_4$ anisotropy pins the vortex into a four-lobed intensity pattern. This makes the vortex resemble a quadrupole in $|\psi|$, but not in its phase structure. A real quadrupole has alternating lobe phases, approximately $(0, \pi, 0, \pi)$, and carries no phase winding. By contrast, the phases of a $D_4$-pinned vortex advance around the core, approximately $(0, \pi/2, \pi, 3\pi/2)$, giving a total winding of $2\pi$ along a closed loop enclosing the center. This nonzero winding requires a complex field and cannot be removed continuously without eliminating the central phase singularity. The vortex event should therefore be interpreted not as a real quadrupole, but as a discrete spacetime vortex pinned by the $D_4$ symmetry of the spacetime.

Experimentally, a generic pulse is unlikely to assemble a higher-order structured event without additional control. The estimated scales can nevertheless be related directly to an optical spacetime crystal operating at visible-to-near-infrared wavelengths. For the representative parameters of Method 1, the spatial modulation period $\Lambda = 1\ \mu m$, modulation depths $\delta_1 = \delta_2 = 0.1$, and a temporal modulation cycle $T_m \approx 2.7\ fs$ for an 800 $nm$ carrier. Under this parameter mapping, one dimensionless unit corresponds to $\Delta x \approx 6.37\ \mu m$ and $\Delta t \approx 17.19\ fs$. The structures spanning 5-10 dimensionless unit therefore correspond to spatial extents of approximately 30-60 $\mu m$ and temporal windows of approximately 85-170 fs. These scales suggest possible preparation routes based on phase-imprinted or parity-imprinted seeds, temporally shaped inputs, or weak dissipative selection, with successful reconstruction requiring the localized core to form within the available preparation window. The present IVP analysis should therefore be regarded as a benchmark for the launch accuracy and preparation window required in future experiments, rather than as an optimized preparation protocol.

## Conclusion

In conclusion, nonlinear photonic spacetime crystals support localized events with internal multipole and vortex structure in the physical $(x, t)$ plane. The action hierarchy identifies their structural cost, the BdG spectrum resolves their deformation channels, and the IVP reconstruction test shows that practical accessibility is governed by finite preparation windows rather than indefinite stability. These results establish structured spacetime events as designable mesoscopic wave objects and provide concrete targets for future optical STC experiments.

## Acknowledgements

**Funding**: This work was supported by the NSFC (No. 2023X0201-417-03) and the start-up funding from ShanghaiTech University.

## Availability of data and materials

The data that support the findings of this study are available from the corresponding authors upon reasonable request.

## Competing interests

The authors declare that they have no competing interests.

## Methods.

### 1. From the nonlinear wave equation to the event-soliton model

We derive the dimensionless scalar equation used in the main text from a one-dimensional nonlinear spacetime photonic medium. The reduction keeps only the ingredients needed for the event-soliton localization: weak periodic modulation in space and time, weak Kerr nonlinearity, a slowly varying envelope inside the complete $\omega k$-gap, and projection onto the degenerate gap-center mode.

We consider a non-magnetic, source-free Kerr medium whose linear permittivity is modulated in both $x$ and $t$. Since temporal interfaces can make the electric field discontinuous, we use the displacement field $D$ as the continuous dynamical variable. The linear part of the constitutive relation is specified by the spacetime-dependent permittivity [8].

$$\epsilon_1(x,t) = \epsilon_r \tilde{\epsilon}(x)\tilde{\epsilon}(t) = \epsilon_r(1+\delta_1\cos(\Omega t))(1+\delta_2\cos(Gx)), \tag{1}$$

where $\epsilon_r$ is the mean relative permittivity, $\delta_{1,2}$ are weak modulation depths, and $\Omega$ and G are the temporal modulation frequency and spatial Bragg wavevector. Inverting the Kerr constitutive relation to leading nonlinear order and substituting it into the source-free Maxwell equations gives the one-dimensional nonlinear wave equation

$$\frac{\partial^2 D}{\partial t^2} = \frac{1}{\mu_0}\frac{\partial^2}{\partial x^2}\left(\frac{D}{\epsilon_0\epsilon_1(x,t)} - \frac{\chi_3 D^3}{\epsilon_0^2\epsilon_1^4(x,t)}\right). \tag{2}$$

The spacetime modulation generates reciprocal scattering between forward and backward components near the mixed $\omega k$-gap. We therefore expand the field in slowly varying envelopes centered at the Bragg wavevector ($\pm$G/2) and half the pump frequency ($\pm\Omega/2$),

$$\tilde{E}(x,t) = A_f e^{i\frac{G}{2}x - i\frac{\Omega}{2}t} + A_b e^{-i\frac{G}{2}x - i\frac{\Omega}{2}t} + \text{c.c.} \tag{3}$$

Substituting into Eq. 2, followed by the slowly varying envelope approximation, gives the coupled mode equations

$$\begin{aligned}
iG\frac{\partial A_f}{\partial x} + i\frac{\Omega}{c^2}\frac{\partial A_f}{\partial t} + \lambda A_f + \kappa_1 A_b^* + \kappa_2 A_b &= \gamma\left(\left|A_f\right|^2 + 2|A_b|^2\right)A_f,\\
-iG\frac{\partial A_b}{\partial x} + i\frac{\Omega}{c^2}\frac{\partial A_b}{\partial t} + \lambda A_b + \kappa_1 A_f^* + \kappa_2 A_f &= \gamma\left(2\left|A_f\right|^2 + |A_b|^2\right)A_b,\\
-iG\frac{\partial A_f^*}{\partial x} - i\frac{\Omega}{c^2}\frac{\partial A_f^*}{\partial t} + \lambda A_f^* + \kappa_1 A_b + \kappa_2 A_b^* &= \gamma\left(\left|A_f\right|^2 + 2|A_b|^2\right)A_f^*,\\
iG\frac{\partial A_b^*}{\partial x} - i\frac{\Omega}{c^2}\frac{\partial A_b^*}{\partial t} + \lambda A_b^* + \kappa_1 A_f + \kappa_2 A_f^* &= \gamma\left(2\left|A_f\right|^2 + |A_b|^2\right)A_b^*,
\end{aligned} \tag{4}$$

where $\lambda = (\Omega^2/c^2 - G^2)/4$ is the detuning, $\kappa_1 = \delta_1 G^2/8$ and $\kappa_2 = \delta_2\Omega^2/8c^2$ define the temporal-coupling and spatial-coupling parameters, respectively, and $\gamma = -3\beta G^2/4$ is the effective Kerr coefficient. At the light-cone modulation point $r = \Omega/(Gc) = 1$, the relevant modes meet at the gap center. Projecting onto the symmetric eigenvector $\chi_d = (1,1,1,1)^T/2$, we write the four-component envelope as $\psi = a(x,t)\chi_d$.Under the resonance condition $\lambda = 0$ and equally-valued spatiotemporal coupling $\kappa_1 = \kappa_2 = \kappa$, the projected dynamics reduce to a single nonlinear envelope equation:

$$\left(1 - \frac{\Omega^2}{4\kappa^2 c^4}\frac{\partial^2}{\partial t^2}\right)\left(1 - \frac{G^2}{4\kappa^2}\frac{\partial^2}{\partial x^2}\right)a - \frac{3\gamma}{8\kappa}|a|^2 a = 0. \quad (5)$$

Here $\kappa$ is the effective coupling induced by the spacetime modulation, and $\gamma$ is the effective Kerr coefficient. Eq. (5) is then brought to the dimensionless form used in the main text by rescaling the spacetime coordinates and the envelope amplitude. The spatial and temporal differential operators are normalized by introducing $x' = \frac{\pi\delta_1}{2\Lambda}x,\ t' = \frac{\pi\delta_2}{2T_m}t$. At the symmetric gap-center point, $\kappa = \kappa_1 = \kappa_2$, with $\kappa_1 = \delta_1 G^2/8$ and $\kappa_2 = \delta_2\Omega^2/(8c^2)$, the scaling becomes $x' = (\delta_1 G/4)x$ and $t' = (\delta_2\Omega/4)t$. Using $G = 2\pi/\Lambda$ (with $\Lambda$ being the spatial modulation period) and the modulation frequency $\Omega = 2\pi/T_m$ (with $T_m$ being the temporal modulation period), this gives

$$x' = \frac{\pi\delta_1}{2\Lambda}x, \quad t' = \frac{\pi\delta_2}{2T_m}t \quad (6)$$

The envelope is rescaled as $a = \sqrt{32\kappa/3\gamma}\,\psi$, so that the nonlinear coefficient is normalized to 4. Dropping the primes, the governing equation becomes

$$(1 - \partial_{tt})(1 - \partial_{xx})\psi - 4|\psi|^2\psi = 0 \quad (7)$$

In this equation, $x$ and $t$ operate entirely symmetrically as coordinates of the event itself, rather than as a propagation coordinate and a retarded time.

This non-dimensionalization relates the dimensionless computational grid to physical spacetime scales. One unit in the dimensionless coordinate, $\Delta x' = 1$ and $\Delta t' = 1$, corresponds to $\Delta x = 2\Lambda/(\pi\delta_1)$ and $\Delta t = 2T_m/(\pi\delta_2)$. For representative nonlinear spacetime crystal parameters, a spatial modulation period of $\Lambda = 1\,\mu$m with a modulation depth of $\delta_1 = 0.1$ and a temporal modulation cycle of $T_m \approx 2.7$ fs for a near-infrared pump at $800$ nm with modulation depth of $\delta_2 = 0.1$, one dimensionless unit corresponds to $\Delta x \approx 6.37\,\mu$m and $\Delta t \approx 17.19$ fs.

Thus, the structured event states in our simulations, which typically span 5-10 dimensionless units, correspond to physical extent of approximately $30$-$60\,\mu$m and temporal window of $85$-$170$ fs. These estimates indicate that the predicted event structures occur on mesoscopic spacetime scales that are compatible with possible spatiotemporal optical characterization techniques.

## 2. Action functional and variational estimates

The mixed operator $(1 - \partial_{tt})(1 - \partial_{xx})$ is invariant under sign reversal and the interchange of $x$ and $t$, but it is not invariant under continuous rotations in the $(x,t)$ plane. This mixed derivative introduces a fourfold angular component, reducing the continuous $O(2)$ symmetry of an isotropic spacetime plane to the discrete $D_4$ spacetime symmetry. This underlying symmetry governs the emergence of real multipole events and complex vortex events.

We treat the governing equation as a boundary-value problem on the full $(x, t)$ plane. For localized fields, assuming $\psi \to 0$ and its derivatives vanish as $|x|, |t| \to \infty$, all boundary terms evaluate to zero upon repeated integration by parts. For reference, the governing dimensionless event-soliton equation, Eq. (7), can be expanded into a fourth-order nonlinear PDE:

$$\psi - \partial_{xx}\psi - \partial_{tt}\psi + \partial_{tt}\partial_{xx}\psi - 4|\psi|^2\psi = 0, \tag{8}$$

We construct an action $S[\psi]$ such that its first variation with respect to the complex conjugate field $\psi^*$ recovers Eq. (8). Integrating by parts under localized boundary conditions, the corresponding spacetime action is given by:

$$S[\psi] = \iint_{-\infty}^{\infty} (|\psi|^2 + |\partial_x\psi|^2 + |\partial_t\psi|^2 + |\partial_x\partial_t\psi|^2 - 2|\psi|^4)\,dx\,dt \tag{9}$$

For complex fields, $\psi$ and $\psi^*$are varied independently. The same action can be written as

$$S[\psi, \psi^*] = \iint_{-\infty}^{\infty} (\psi\psi^* + (\partial_x\psi)(\partial_x\psi^*) + (\partial_t\psi)(\partial_t\psi^*) + (\partial_x\partial_t\psi)(\partial_x\partial_t\psi^*) - 2(\psi\psi^*)^2)\,dx\,dt \tag{10}$$

The first variation with respect to $\psi^*$ is reduced by moving all derivatives from $\delta\psi^*$ onto $\psi$. Because the boundary terms vanish, one obtains

$$\delta S = \iint_{-\infty}^{\infty} dx\,dt\,\delta\psi^*[\psi - \partial_{xx}\psi - \partial_{tt}\psi + \partial_{tt}\partial_{xx}\psi - 4|\psi|^2\psi] \tag{11}$$

Stationarity, requiring $\delta S = 0$ for arbitrary $\delta\psi^*$, gives the expanded Euler-Lagrange equation, which factorizes back to $(1 - \partial_{tt})(1 - \partial_{xx})\psi - 4|\psi|^2\psi = 0$. (Eq. 7)

The action values presented in Fig. 1 should therefore be interpreted as relative costs for different stationary events, rather than dynamical energies. This scalar measure allows monopole, dipole, vortex, quadrupole and octupole configurations to be compared within the same spacetime functional landscape.

To obtain analytic estimates, we use trial functions that enforce the desired spacetime symmetry. For example, a real $x$-odd dipole (space dipole) is parameterized by

$$\psi_{dipole}(x, t) = A \cdot x \cdot exp\big(-\alpha(x^2 + t^2)\big) \tag{12}$$

Here, $A$ is the amplitude and $\alpha$ is the inverse squared width. Substitution into the functional $S$ reduces the infinite-dimensional variational problem to a finite-dimensional action $S(A, \alpha)$; anisotropic extensions can use separate widths in $x$ and $t$. The variational parameters are fixed by the stationarity conditions

$$\frac{\partial S}{\partial A} = 0, \quad \frac{\partial S}{\partial \alpha} = 0 \tag{13}$$

This identical construction provides symmetry-adapted trial functions for temporal dipole, quadrupole, vortex and octupole events:

$$\begin{gathered}\psi_{dipole'}(x,t) = A \cdot t \cdot exp\left(-\alpha(x^2+t^2)\right) \\ \psi_{quad}(x,t) = A \cdot xt \cdot exp\left(-\alpha(x^2+t^2)\right) \\ \psi_{vortex}(x,t) = A \cdot (x \pm it) \cdot exp\left(-\alpha(x^2+t^2)\right) \\ \psi_{oct}(x,t) = A \cdot xt(x^2-t^2) \cdot exp\left(-\alpha(x^2+t^2)\right)\end{gathered} \tag{14}$$

These ansatzes are deliberately restrictive. They impose the target nodes lines or phase winding, but freeze the radial profile and approximate the monopole $sech(x)sech(t)$ by a Gaussian envelope. Consequently, we utilize these variational forms strictly as initial estimates and action-scale guides; the exact event profiles are obtained via the symmetry-projected Petviashvili relaxation method detailed in Method 3.

### 3. Symmetry-projected Petviashvili relaxation

The variational estimates in Method 2 provide symmetry-compatible initial guesses, but the final event profiles are obtained from the full boundary-value problem. We solve the stationary equation in the generic form (Eq. 5), where the linear and nonlinear parts for the event-soliton model are $L_{xt} = (1-\partial_{tt})(1-\partial_{xx})$ and $\mathcal{N}[\psi] = 4|\psi|^2\psi$. The computation is performed on a square pseudospectral grid large enough that the profile decays before reaching the boundary.

$$L_{xt}\psi(x,t) = \mathcal{N}[\psi(x,t)], \tag{15}$$

Fourier discretization transforms the linear operator into an algebraic multiplier and avoids finite-difference errors in the mixed derivative. In spectral form, the equation is

$$L(k_x,k_t)\tilde{\psi}(k_x,k_t) = \mathcal{F}\{\mathcal{N}[\psi(x,t)]\} \tag{16}$$

A direct fixed-point update of the spectral equation is unstable for self-focusing localized states, because the iteration either collapses to zero or grows without bound. We therefore employ a Petviashvili iteration, governed by the update rule

$$\tilde{\psi}^{(m+1)} = \frac{M^{\gamma} \cdot \mathcal{F}\{\mathcal{N}[\psi^{(m)}]\}}{L(k_x,k_t)} \tag{17}$$

where $\gamma = 1.5$ for a cubic Kerr nonlinearity [22], and the stabilization factor $M$ balances the linear and nonlinear contributions:

$$M = \frac{\langle\tilde{\psi}^{(m)}|L(k_x,k_t)|\tilde{\psi}^{(m)}\rangle}{\langle\tilde{\psi}^{(m)}|\mathcal{F}^{-1}\{\mathcal{N}[\psi^{(m)}]\}\rangle} = \frac{\int L(k_x,k_t)\left|\tilde{\psi}^{(m)}(k_x,k_t)\right|^2 dk_x dk_t}{\int \tilde{\psi}^{(m)*}(k_x,k_t)\,\mathcal{F}\{\mathcal{N}[\psi^{(m)}]\}\,dk_x dk_t} \tag{18}$$

Higher-order events are inherently saddle points; thus, unconstrained relaxation unavoidably collapses to the fundamental monopole. To stabilize the iteration, we apply strict parity or phase-winding projections after each inverse Fourier transform. For example, an octupole event is isolated by imposing odd parities in both dimensions ($\psi(x,t) \rightarrow [\psi(x,t) - \psi(-x,t)]/2$ and $\psi(x,t) \rightarrow [\psi(x,t) - \psi(x,-t)]/2$) combined with diagonal antisymmetry ($\psi(x,t) \rightarrow [\psi(x,t) - \psi(t,x)]/2$). For vortex events, the configuration is dynamically projected onto the $m = 1$ complex azimuthal phase sector to preserve the topological winding.

4. **BdG spectrum analysis**

The second variation of the spacetime action identifies localized deformation channels, zero modes, and unstable directions on the saddle manifold within the full function space. For each stationary event profile, we compute the second variation of the spacetime action given in Eq. (9). Subjecting a complex stationary profile $\psi_0$ to a perturbation $\psi = \psi_0 + \delta\psi$ and expanding the action to the leading quadratic order, the perturbation action $S_2$ is obtained as:

$$S_2 = \iint \left[\delta\psi^* \hat{L} \delta\psi - 8|\psi_0|^2 \delta\psi^* \delta\psi - 2\psi_0^2 (\delta\psi^*)^2 - 2(\psi_0^*)^2 (\delta\psi)^2\right] dxdt \tag{19}$$

Equivalently, the quadratic form can be expressed in the Nambu basis $\Phi = (\delta\psi, \delta\psi^*)^T$ as:

$$S_2 = \frac{1}{2} \iint (\delta\psi^* \quad \delta\psi) \begin{pmatrix} \mathcal{H}_{11} & \mathcal{H}_{12} \\ \mathcal{H}_{21} & \mathcal{H}_{22} \end{pmatrix} (\delta\psi \quad \delta\psi^*)^T dxdt \tag{20}$$

Comparison of the quadratic terms gives the complex-field BdG operator:

$$\mathcal{H} = \begin{pmatrix} L_{xt} - 8|\psi_0|^2 & -4\psi_0^2 \\ -4(\psi_0^*)^2 & L_{xt} - 8|\psi_0|^2 \end{pmatrix} \tag{21}$$

where $L_{xt} = (1 - \partial_{tt})(1 - \partial_{xx})$. For a vortex event with $\psi_0 \propto e^{im\theta}$, the off-diagonal terms contain the doubled phase of the base state, $\mathcal{H}_{12} \propto \psi_0^2 \propto e^{i2m\theta}$ and $\mathcal{H}_{21} \propto (\psi_0^*)^2 \propto e^{-i2m\theta}$. Thus, vortex bound modes couple amplitude and phase perturbations through the winding of the underlying event. This constitutes the fundamental spectral distinction between a complex vortex event and a real quadrupole that possesses a visually similar four-sector intensity pattern.

For real multipole events, the field lacks an independent phase channel. Writing $\psi = \phi_0 + U$ gives the real action

$$S^{(2)} = \iint U\left[\hat{L} - 12\phi_0^2\right]U \; dxdt \tag{22}$$

alongside the scalar BdG operator:

$$\mathcal{H}_{real} = L_{xt} - 12\phi_0^2 \tag{23}$$

The BdG operators are discretized with the same pseudospectral grid used for the stationary profiles. Because only the low-lying localized spectrum is needed, we compute selected eigenpairs using shift-invert Arnoldi iteration rather than full diagonalization. For a target shift $\sigma$, the transformed eigenproblem is formulated as:

$$(\mathcal{H} - \sigma \mathrm{I})^{-1}\psi = \frac{1}{\lambda - \sigma}\psi \tag{24}$$

Eigenvalues $\lambda$ near $\sigma$ are mapped to large transformed eigenvalues $\mu = 1/(\lambda - \sigma)$, enabling the Arnoldi iteration to efficiently resolve modes inside or near the gap. The physical eigenvalue is recovered via $\lambda = \sigma + 1/\mu$. Modes are classified as bound or continuum-like by their eigenvalue position relative to the continuum threshold and by their spatial localization. Translational and global phase zero modes, when present, are used as checks on the discretization and boundary size.

Real multipoles are analyzed with the scalar BdG operator $\mathcal{H}_{real}$, while complex vortex events require the full BdG block structure. This is why the vortex and quadrupole spectra shown in Fig. 2 have different internal fingerprints even when their intensities look similar.

## 5. **Initial-value evolution and finite excitation windows**

Initial-value problem (IVP) simulations act as an independent dynamical robustness check, verifying whether a localized time slice can reconstruct the target event before diverging along unstable directions. Using the high-fidelity target profile $\psi(x,t)$ directly from the Petviashvili solver, we initialize the system at a negative-time slice $t_0$ preceding the maximum core interaction.

To initialize the IVP and avoid finite-difference truncation errors that artificially excite evanescent branches, we accurately compute the initial velocity field spectrally from the full boundary-value profile:

$$\partial_t \psi_{theory}(x,t) = \mathcal{F}_{2D}^{-1}\left[ i k_t \mathcal{F}_{2D}\left[\psi_{theory}(x,t)\right]\right] \tag{25}$$

The slice $\psi(x,t_0)$ and the spectral derivative $\partial_t \psi(x,t_0)$ are then used as the initial displacement and velocity.

For time stepping we introduce $U = (1 - \partial_{xx})\psi$ and the pseudospectral inverse $L_{inv} = (1 - \partial_{xx})^{-1}$. The governing equation is then written as a first-order real system. Because the Kerr term contains the non-holomorphic quantity $|\psi|^2\psi$, the complex field is split into real and imaginary parts. With $U = U_R + iU_I$ and $V = \partial_t U = V_R + iV_I$, the evolution equations are

$$\begin{gathered}
\partial_t U_R = V_R \\
\partial_t U_I = V_I \\
\partial_t V_R = U_R - 4\left(\left(\hat{L}_{inv}U_R\right)^2 + \left(\hat{L}_{inv}U_I\right)^2\right)\left(\hat{L}_{inv}U_R\right) \\
\partial_t V_I = U_I - 4\left(\left(\hat{L}_{inv}U_R\right)^2 + \left(\hat{L}_{inv}U_I\right)^2\right)\left(\hat{L}_{inv}U_I\right)
\end{gathered} \tag{26}$$

The initial data for $U$ and $V$ are obtained by applying the forward linear operator $(1 - \partial_{xx})$ to the field and velocity slices. Time evolution is performed with an explicit fourth-order Runge-Kutta scheme and pseudospectral evaluation of the spatial operators. The step size and computational window are checked by comparing the reconstructed event core under refinement. The codes are provided in the SM file.

Because the event profiles are saddle-type homoclinic objects, exponentially growing components are unavoidable over long times. We therefore analyze only a finite preparation window: the physically relevant interval is the approach to the target core and the early departure from it, before numerical and physical unstable directions overwhelm the reconstruction.

## Reference

[1] B.A. Malomed, (INVITED) Vortex solitons: Old results and new perspectives, Physica D 399 (2019) 108–137. https://doi.org/10.1016/j.physd.2019.04.009.

[2] C. Caloz, Z.L. Deck-Leger, Spacetime Metamaterials-Part I: General Concepts, IEEE Trans. Antennas Propag. 68 (2020) 1569–1582. https://doi.org/10.1109/TAP.2019.2944225.

[3] C. Caloz, Z.L. Deck-Leger, Spacetime Metamaterials-Part II: Theory and Applications, IEEE Trans. Antennas Propag. 68 (2020) 1583–1598. https://doi.org/10.1109/TAP.2019.2944216.

[4] E. Galiffi, R. Tirole, S. Yin, H. Li, S. Vezzoli, P.A. Huidobro, M.G. Silveirinha, R. Sapienza, A. Alù, J.B. Pendry, Photonics of time-varying media, Advanced Photonics 4 (2022). https://doi.org/10.1117/1.AP.4.1.014002.

[5] F.R. Morgenthaler, Velocity Modulation of Electromagnetic Waves, IEEE Trans. Microw. Theory Tech. 6 (1958) 167–172. https://doi.org/10.1109/TMTT.1958.1124533.

[6] D. Holberg, K. Kunz, Parametric properties of fields in a slab of time-varying permittivity, IEEE Trans. Antennas Propag. 14 (1966) 183–194. https://doi.org/10.1109/TAP.1966.1138637.

[7] E. Lustig, Y. Sharabi, M. Segev, Topological aspects of photonic time crystals, Optica 5 (2018) 1390. https://doi.org/10.1364/optica.5.001390.

[8] L. Zhang, Z. Fan, Y. Pan, Event soliton formation in mixed energy-momentum gaps of nonlinear spacetime crystals, Phys. Rev. Res. 8 (2026). https://doi.org/10.1103/q226-dkk4.

[9] Z. Zhu, B. Huang, S. Xu, J. Chen, Y. Meng, Z. Zhu, X. Xi, Z. Gao, Spatiotemporal topological phase transitions in photonic spacetime crystals, n.d.

[10] Y. Sharabi, A. Dikopoltsev, E. Lustig, Y. Lumer, M. Segev, Spatiotemporal photonic crystals, Optica 9 (2022) 585. https://doi.org/10.1364/OPTICA.455672.

[11] J. Feis, S. Weidemann, T. Sheppard, H.M. Price, A. Szameit, Space-time-topological events in photonic quantum walks, Nat. Photonics 19 (2025) 518–525. https://doi.org/10.1038/s41566-025-01653-w.

[12] Y. Pan, M.I. Cohen, M. Segev, Superluminal k -Gap Solitons in Nonlinear Photonic Time Crystals, Phys. Rev. Lett. 130 (2023). https://doi.org/10.1103/PhysRevLett.130.233801.

[13] B.J. Eggleton, R.E. Slusher, C.M. de Sterke, P.A. Krug, J.E. Sipe, Bragg Grating Solitons, Phys. Rev. Lett. 76 (1996) 1627–1630. https://doi.org/10.1103/PhysRevLett.76.1627.

[14] D.N. Christodoulides, R.I. Joseph, Slow Bragg solitons in nonlinear periodic structures, Phys. Rev. Lett. 62 (1989) 1746–1749. https://doi.org/10.1103/PhysRevLett.62.1746.

[15] W. Chen, D.L. Mills, Gap solitons and the nonlinear optical response of superlattices, Phys. Rev. Lett. 58 (1987) 160–163. https://doi.org/10.1103/PhysRevLett.58.160.

[16] H.A. Haus, W.S. Wong, Solitons in optical communications, Rev. Mod. Phys. 68 (1996) 423–444. https://doi.org/10.1103/RevModPhys.68.423.

[17] Y. Silberberg, Collapse of optical pulses, 1990. https://doi.org/10.1364/OA_License_v1#VOR.

[18] X. Liu, L.J. Qian, F.W. Wise, Generation of Optical Spatiotemporal Solitons, Phys. Rev. Lett. 82 (1999) 4631–4634. https://doi.org/10.1103/PhysRevLett.82.4631.

[19] B.A. Malomed, D. Mihalache, F. Wise, L. Torner, Spatiotemporal optical solitons, Journal of Optics B: Quantum and Semiclassical Optics 7 (2005). https://doi.org/10.1088/1464-4266/7/5/R02.

[20] H.E. Kondakci, A.F. Abouraddy, Diffraction-free space-time light sheets, Nat. Photonics 11 (2017) 733–740. https://doi.org/10.1038/s41566-017-0028-9.
[21] M. Yessenov, L.A. Hall, K.L. Schepler, A.F. Abouraddy, Space-time wave packets, Adv. Opt. Photonics 14 (2022) 455. https://doi.org/10.1364/AOP.450016.
[22] D.E. Pelinovsky, Y.A. Stepanyants, Convergence of petviashvili's iteration method for numerical approximation of stationary solutions of nonlinear wave equations, SIAM J. Numer. Anal. 42 (2004) 1110–1127. https://doi.org/10.1137/S0036142902414232.

# Figures

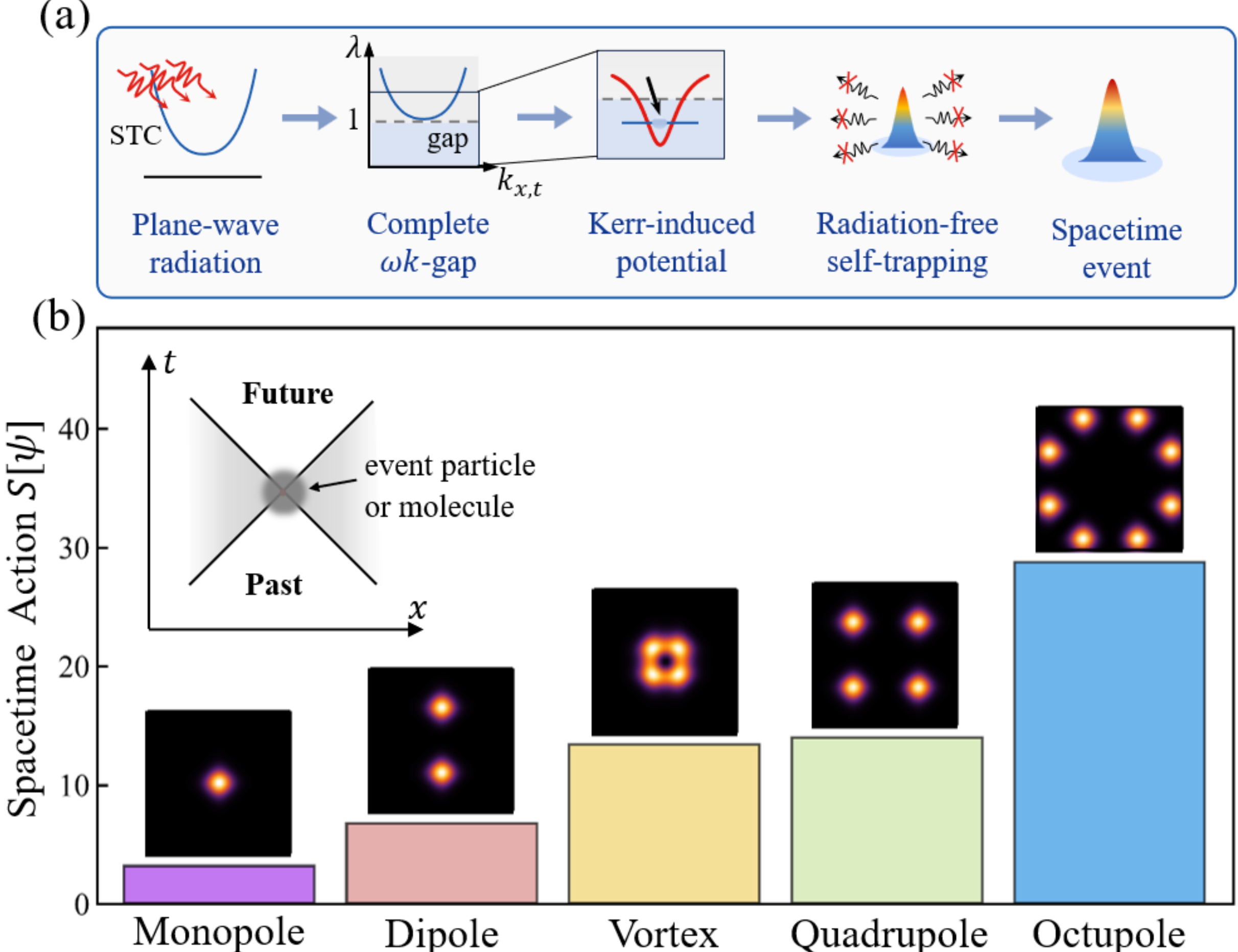

**Figure 1: From plane-wave excitation to event particles and event molecules.** (a) In the nonlinear spacetime crystal, Kerr self-action lowers the effective eigenvalue of a continuum excitation into a semi-infinite gap, suppressing radiative leakage and yielding a localized event soliton in the $\omega k$-gap. (b) The spacetime-action hierarchy interprets the resulting localized solutions as physical events with internal structure. Insets contrast a mathematical event, represented by a zero-dimensional spacetime point, with a physical event wave packet of finite spatial and temporal extent. The bars order the stationary Lyapunov action $S[\psi]$ of monopole, dipole, vortex, quadrupole and octupole event solitons; image insets show representative localized profiles. The monopole is the elementary event particle, whereas the higher-order states are event molecules assembled from multiple spacetime lobes, nodal lines or a phase singularity. The ordering $S_{monopole} < S_{dipole} < S_{vortex} < S_{quadrupole} < S_{octupole}$ quantifies the action cost of internal spacetime structure. The lower action of the vortex relative to the quadrupole indicates that a point-like phase singularity is less costly than extended real nodal-line constraints.

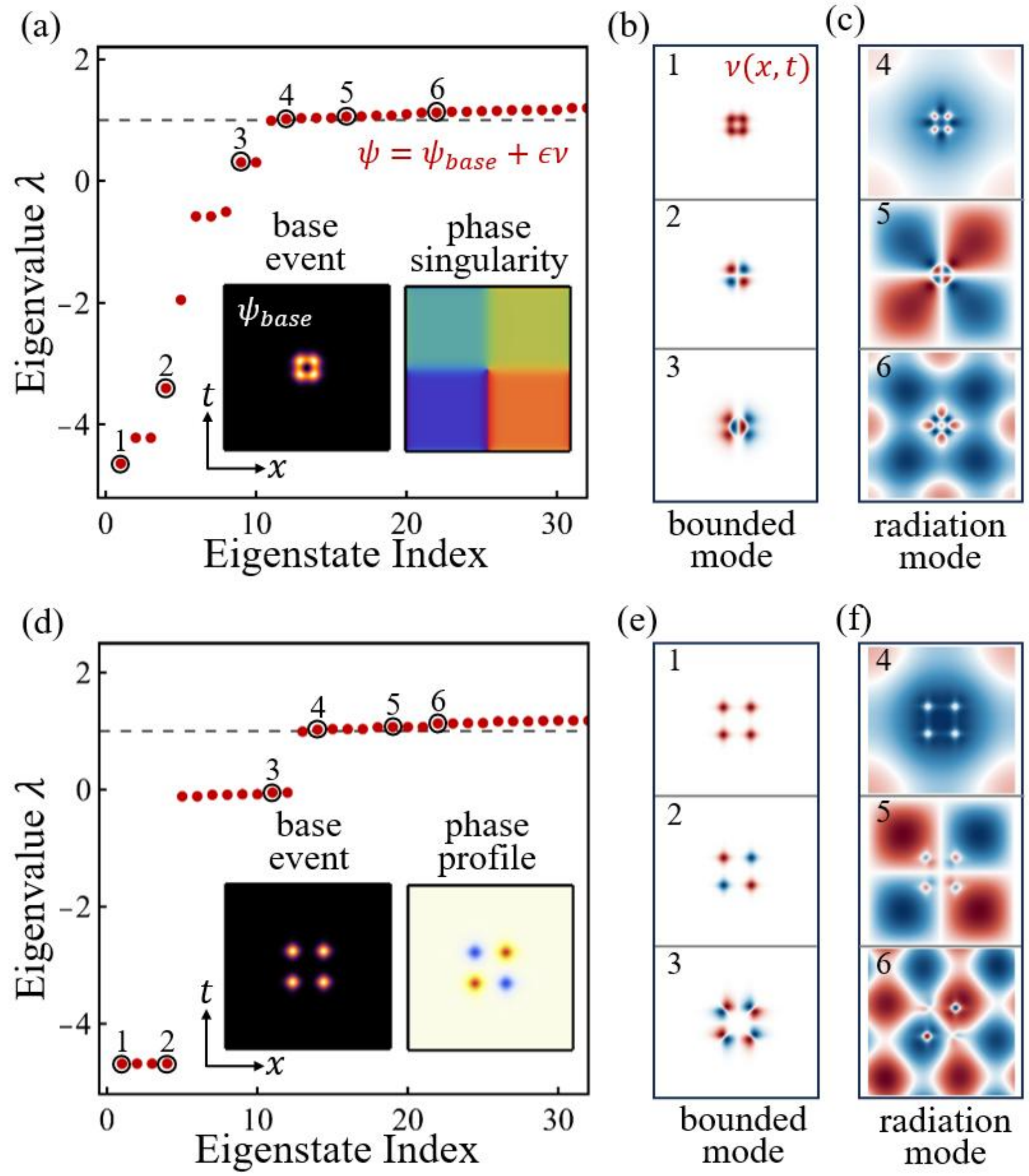

**Figure 2: BdG spectra as fingerprints of vortex and quadrupole event molecules.** (a) Eigenvalue spectrum of the vortex event; the dashed line marks $\lambda = 1$, the continuum threshold, and the insets show the intensity and four-sector phase winding of the base event. (b, c) Representative modes associated with the numbered eigenstates in (a): bound modes below threshold are localized at the vortex core, whereas continuum modes extend across the spacetime domain. (d) Eigenvalue spectrum of the quadrupole event, with insets showing its four-lobed real parity structure. (e, f) Corresponding bound and continuum modes of the quadrupole. Although the vortex and quadrupole can display similar four-sector intensity patterns, their spectra separate a phase-winding point defect from real nodal-line constraints.

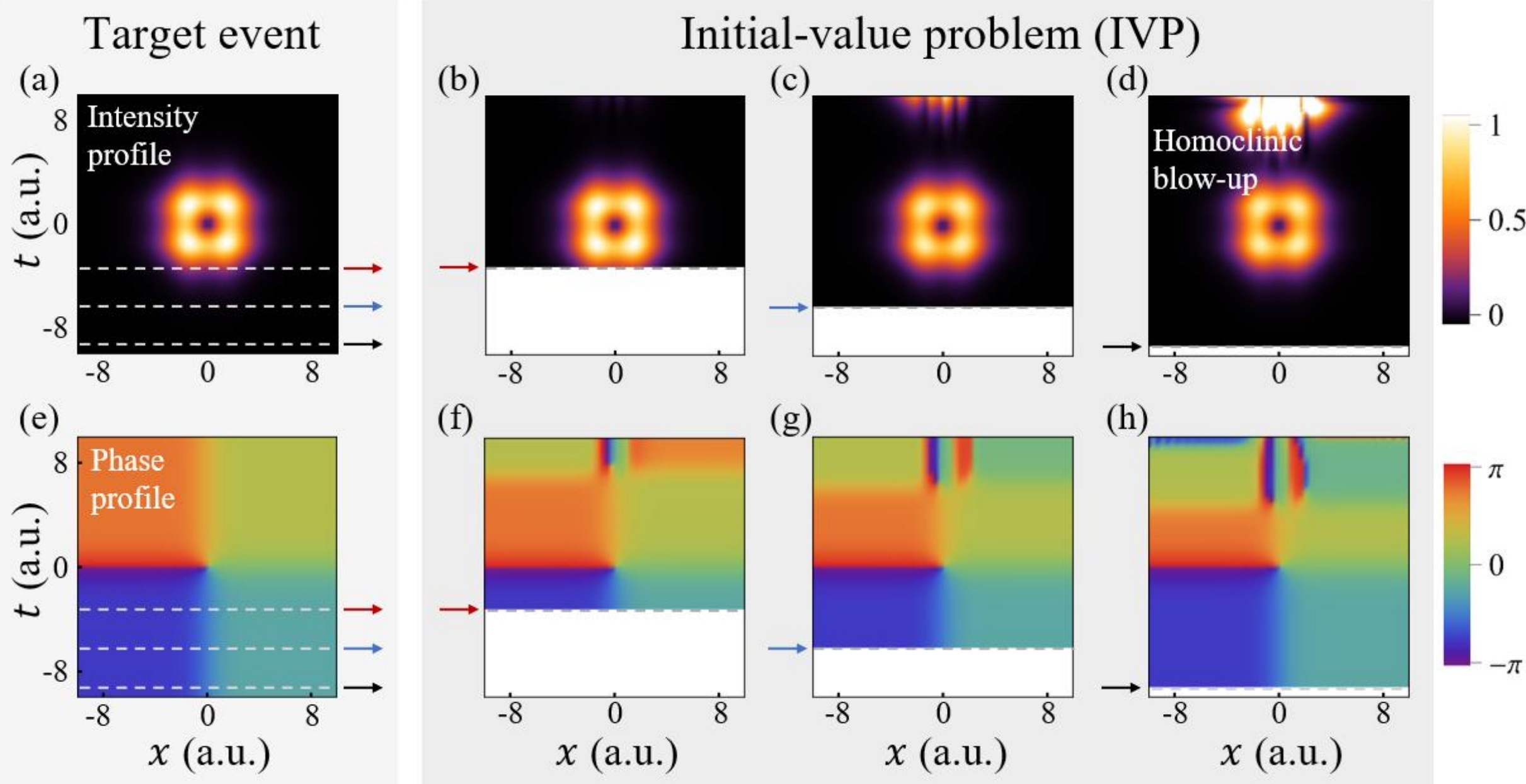

Figure 3: Finite spacetime-window reconstruction of a vortex event. (a, e) Target spacetime vortex event, showing localized intensity and the four-sector phase texture associated with the $m = 1$ winding. Dashed lines mark three negative-time slices used as IVP initial conditions. (b-d, f-h) IVP evolutions launched from these slices. A sufficiently late slice reconstructs both the vortex core and phase winding near ($t \approx 0$), providing an independent dynamical check of the stationary solution. Earlier slices require longer pre-assembly evolution, allowing unstable components to grow; the trajectory then exits the local neighborhood through homoclinic blow-up and phase dislocations. Successful preparation therefore requires an initial slice that remains close enough to the target state for the core to reconstruct.

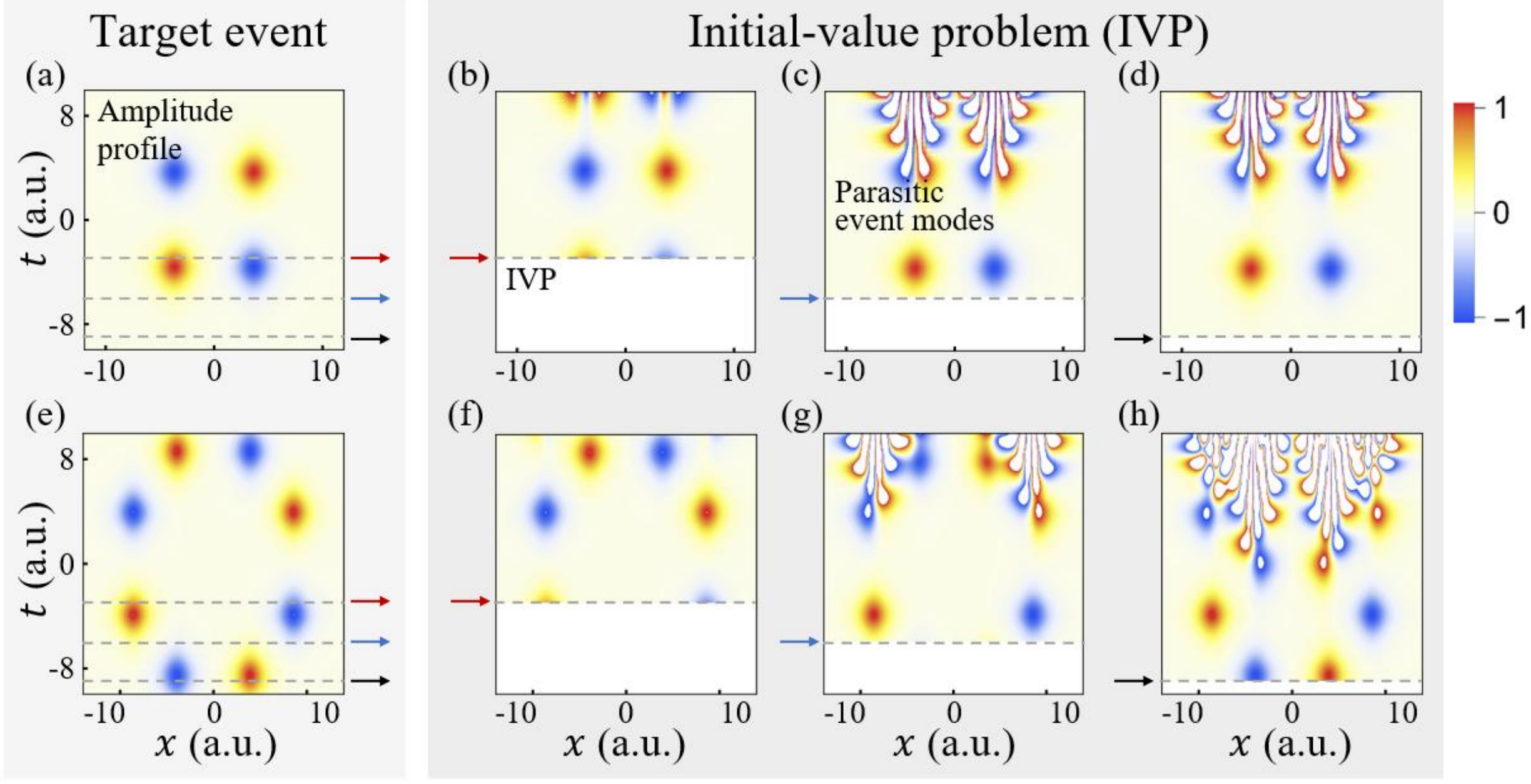

Figure 4: Order-dependent fragility in the IVP reconstruction of real multipole events. (a) Target quadrupole event with four-lobed real parity structure; dashed lines mark three initial slices. (b-d) IVP evolutions from progressively earlier slices transiently recover the quadrupole near the interaction region, but the nodal structure then breaks into fingering patterns and homoclinic blow-up. (e) Target octupole event. (f-h) Corresponding octupole evolutions show a narrower finite preparation window and faster structural collapse, because additional lobes and sign changes amplify small mismatches into filamentary patterns. The trend connects the IVP dynamics to the action hierarchy and to the denser set of BdG deformation channels in higher-order event states.